\documentclass[11pt,letterpaper,notoc,nohyper]{JHEP}
\usepackage{amsmath}
\usepackage{cite}
\newcommand{\SM}{ Standard Model}
\DeclareMathOperator{\tr}{\rm tr}
\DeclareMathOperator{\diag}{\rm diag}
\newcommand\openone{\leavevmode\hbox{\small1\normalsize\kern-.33em1}}
\providecommand{\abs}[1]{\lvert#1\rvert}
\def\CL{{\cal L}}

\def\CO{{\cal O}}
\def\CL{{\cal L}}

\def\one{{\bf 1}}

\def\be{\begin{equation}}
\def\ee{\end{equation}}
\def\bea{\begin{eqnarray}}
\def\eea{\end{eqnarray}}

\newcommand{\gsim}{ \mathop{}_{\textstyle \sim}^{\textstyle >} }

\newcommand\beq{\begin{eqnarray}}
\newcommand\eeq{\end{eqnarray}}

\title{A Composite Little Higgs Model}
\author{Emanuel Katz, Jaeyong Lee, Ann E. Nelson \\ Department of
  Physics, Box 1560, University of Washington, Seattle, WA 98195-1560
  \\ email: \email{amikatz@fermi.phys.washington.edu},
  \email{anelson@phys.washington.edu}, \email{jaeyong@u.washington.edu}}
\author{Devin G. E. Walker\\ Jefferson Laboratory of Physics, Harvard
  University, Cambridge, MA 02138 \\ email:
  \email{walker@lamb.physics.harvard.edu}} 
  
  \preprint{UW/PT-03-13,\  HUTP-03/A050} 
 \abstract{ We describe a natural UV complete theory with  a
composite little Higgs.  Below a TeV we have the minimal Standard
Model with a light Higgs, and an extra neutral scalar.  At the TeV
scale there are  additional scalars, gauge bosons, and
vector-like charge 2/3 quarks, whose couplings to the Higgs greatly
reduce the  UV sensitivity of  the Higgs potential. Stabilization of
the Higgs mass squared parameter, without finetuning, occurs due to a
softly broken shift symmetry---the Higgs is a pseudo Nambu-Goldstone
boson.  Above the 10 TeV scale the theory has new strongly coupled
interactions.  A  perturbatively
renormalizable UV completion, with softly broken supersymmetry at 10
TeV is explicitly worked out.  Our theory contains new particles which
are odd under an exact ``dark matter parity", $(-1)^{(2S+3B+L)}$. We argue that such a parity is likely to be a feature of many theories of new TeV scale physics.  The
lightest  parity odd  particle,  or ``LPOP", is  most likely a neutral
fermion,  and may make a good dark matter candidate, with similar
experimental signatures to the neutralino of the MSSM.  We give a
general effective field theory analysis of the calculation of
corrections to precision electroweak observables.}
\begin{document}

\section{Introduction }
\label{intro}
A new mechanism for electroweak symmetry breaking, dubbed the ``little
Higgs'' \cite{Arkani-Hamed:2001nc}, was recently discovered via
dimensional deconstruction
\cite{Arkani-Hamed:2001ca,Hill:2000mu}. This mechanism  has since been
realized in various simple nonlinear sigma models
\cite{Arkani-Hamed:2002qy,Arkani-Hamed:2002qx,Low:2002ws,Kaplan:2003uc,Chang:2003un,Skiba:2003yf,Chang:2003zn}.
In these theories, the Higgs is a pseudo Nambu-Goldstone boson, whose
mass squared is protected against large radiative corrections by
approximate nonlinearly realized global symmetries. These global
symmetries are   explicitly broken.  However the symmetry breaking is
sufficiently soft that  the Higgs mass squared is at most
logarithmically sensitive to the cutoff at one loop, provided the
symmetry breaking terms  satisfy a mild criterion: {\it no {\bf
single}  term in the Lagrangian breaks {\bf all} the  symmetry which
is protecting the Higgs mass}. A collection of terms gives the Higgs
its symmetry-breaking potential, gauge couplings, and Yukawa
couplings. Such symmetry breaking may be thought of as being
``nonlocal in theory space''. Several new,
weakly coupled particles are found at and possibly below a few TeV,
which cancel the leading quadratic divergences in the Higgs mass in a
manner reminiscent of softly broken supersymmetry. However, unlike
supersymmetry, the cancellations occur between particles of the same
statistics. The spectrum and phenomenology of little Higgs theories
have been discussed in
refs. \cite{Arkani-Hamed:2002pa,Gregoire:2002ra,Burdman:2002ns,Huo:2003vd,Han:2003gf,Choudhury:2003ut,Dib:2003zj,Han:2003wu,Perelstein:2003wd}.

In some of the simplest little Higgs theories,   corrections to
precision electroweak observables  are comparable in size to one-loop
\SM\ effects over much of the parameter
space\cite{Csaki:2002qg,Hewett:2002px,Chivukula:2002ww,Han:2003wu,Gregoire:2003kr,
Csaki:2003si},
providing important constraints. It is, however,  straightforward to
find natural, simple,  and experimentally viable little Higgs theories
where the corrections are much smaller
\cite{Han:2003wu,Chang:2003zn,Gregoire:2003kr,
Csaki:2003si,Chang:2003un,Cheng:2003ju}. This will be explicitly demonstrated in
the  model presented in this paper.

A pressing issue   is to situate the little Higgs in a more complete
theory with a higher cutoff.  This is necessary in order to address in
a compelling way the  phenomenology of flavor changing neutral
currents\cite{Lane:2002pe,Chivukula:2002ww}, which is sensitive to
physics beyond the 10 TeV scale.  Further, unitarity constraints from little Higgs
theories make a completion mandatory\cite{Chang:2003vs}. In this paper we 
embed a slightly
altered  version of the ``littlest'' \cite{Arkani-Hamed:2002qy} Higgs
model into a UV complete theory.  The model is experimentally viable,
with acceptable precision electroweak corrections and no fine tuning.
The Higgs is a composite of fermions interacting via strong dynamics
at the 10 TeV scale. The  quark and lepton masses can be generated
without excessive flavor changing neutral currents. These interactions
result from the Yukawa couplings of very heavy scalars, and can naturally
arise in a  theory with supersymmetry softly broken at 10 TeV. We
stress that  many other UV completions of little Higgs theories are
possible, and the present model should by no means be taken as
canonical. We simply wish to explore an explicit model thoroughly  in
order to better understand the theoretical and experimental
implications of the little Higgs.

This model illustrates several advantages of composite little Higgs
models as compared with the traditional approaches to electroweak
symmetry breaking. In contrast to the Minimal Supersymmetric model
(MSSM), there are no problems with fine tuning, lepton flavor
violation, CP violation, or flavor changing neutral currents. The
natural expectation for the masses of the new particles is well above
the weak scale. The chief drawback of the model  relative to the MSSM
is the lack of a prediction for the weak angle.

This UV completion of the littlest Higgs model is quite similar to the
Georgi-Kaplan Composite
Higgs\cite{Kaplan:1984fs,Kaplan:1984sm,Georgi:1984af,Georgi:1984ef,Dugan:1985hq}. However
unlike the Georgi-Kaplan models, in our model the hierarchy between
the compositeness scale and the electroweak scale is not due to  fine
tuning of  parameters.

We will describe this model from the bottom up, as a sequence of
natural effective field theories.  We start in section \ref{littlest}
with a description of the most important physics below 10 TeV. In
section \ref{comp} we describe the strong dynamics which could lead to
such an effective theory, and give a complete, renormalizable theory
of the arbitrarily short distance physics. In section \ref{dark} we
discuss ``dark matter parity''. This is a discrete symmetry which is
an automatic consequence of baryon and lepton number conservation. In
supersymmetry theories this symmetry is known as``R-parity", but the
existence of such a symmetry is well motivated  without supersymmetry.  In
section \ref{precision} we give an effective field theory analysis of
the corrections to precision electroweak observables and the resulting
bounds on the model. Section \ref{flavor} addresses flavor physics and
flavor changing neutral currents.

\section{The $SU(5)/SO(5)$
little Higgs Model}
\label{littlest}
The simplest of the little Higgs models, the ``littlest Higgs'',  is
a nonlinear sigma model whose target space is  the coset space
$SU(5)/SO(5)$ \cite{Arkani-Hamed:2002qy}.  This theory describes the
low energy interactions of 14 Nambu-Goldstone bosons (NGBs), with
decay constant $f\sim1-2$~ TeV. The cutoff of this theory can be as
high as $4\pi f\sim$10~TeV, where the  model becomes strongly coupled.
The SU(5) symmetry is explicitly broken by $\CO(1)$ gauge interactions
and fermion couplings,  leading to masses for most of the NGBs of
order $ g f$, while others get eaten by  gauge bosons whose masses are
also of order $ g f$. A special subset of the NGBs, however, do not
receive masses to leading order in the symmetry breaking terms, and
are about a factor of $g^2/4\pi$ lighter than $f$. In the minimal
model of ref. \cite{Arkani-Hamed:2002qy}, this subset consisted only
of a single Higgs doublet,  dubbed the little Higgs. In the present
model a neutral scalar may also remain light. At the TeV scale a small
number of additional scalars, vector bosons and quarks cancel the one
loop quadratic divergence in the Higgs mass without fine tuning or
supersymmetry.

In order to make this paper self-contained, we describe the effective
nonlinear sigma model in some detail here.  We will describe the
breaking as arising from a vacuum expectation value for a $5\times 5$
symmetric matrix $\Phi$, which transforms as $\Phi \to V \Phi V^T$
under $SU(5)$.  As we will see in the next section, it is
straightforward to construct an explicit composite Higgs model where
this $\Phi$ corresponds to a fermion bilinear. A vacuum expectation
value for $\Phi$ proportional to the unit matrix  breaks $SU(5) \to
SO(5)$. For later convenience, we use an equivalent basis where the
vacuum expectation value for the symmetric tensor points in the
$\Sigma_0$ direction where $\Sigma_0$ is
\begin{equation}
  \label{sigma}
  \Sigma_0=
  \begin{pmatrix} 
    \quad  & \quad & \openone \\ \quad & 1 &\quad \\ \openone
    &\quad&\quad
  \end{pmatrix}\ .
\end{equation}
The unbroken $SO(5)$ generators satisfy
\begin{align}
  T_a \Sigma_0 + \Sigma_0 T_a^T = 0 \intertext{while the broken
generators obey}
  \label{brokengen}
  X_a \Sigma_0 - \Sigma_0 X_a^T = 0\ .
\end{align}
The Nambu-Goldstone bosons are fluctuations about this background in
the broken directions $\Pi \equiv \pi^a X^a$, and can be parameterized
by the non-linear sigma model field
\begin{equation}
  \Sigma(x) = e^{i \Pi/f} \Sigma_0 e^{i \Pi^T/f} =  e^{2 i \Pi/f}
  \Sigma_0 \ .
\end{equation}

We now introduce the gauge and Yukawa interactions which explicitly
break the global symmetry. In ref. \cite{Arkani-Hamed:2002qy}, these
were chosen to ensure an  $SU(3)$ global symmetry under which the
little Higgs transformed nonlinearly, in the limit where any of the
couplings were turned off.   This required embedding the \SM\
$SU(2)_w\times U(1)_y$ gauge interaction  into an $[SU(2)\otimes
U(1)]^2$ gauge group, which was spontaneously broken to
$SU(2)_w\otimes U(1)_y$ at the scale $f$. The extra U(1) and a factor
of (1/5) in the charge matrix led to a rather light $Z'$,  which was
constrained by Tevatron and  precision electroweak corrections
\cite{Csaki:2002qg,Hewett:2002px,Han:2003wu}.  Due to the small size
of the weak angle, we can eliminate the $Z'$ and the associated
constraints without increasing the finetuning of the theory. With a 10
TeV cutoff, naturalness does not require  cancellation of its
quadratically cutoff sensitive contribution to the Higgs mass squared
from weak hypercharge gauge  interactions.  We therefore only
introduce a single U(1). This simplification will make cancellation of
gauge anomalies easy in the underlying composite model, as well
as relaxing experimental constraints.  We thus weakly gauge an $
SU(2)^2 \times U(1)_y$ subgroup of the $SU(5)$ global symmetry. The
generators of the $SU(2)$'s are embedded into $SU(5)$ as
\begin{align}
  Q_1^a &=
  \begin{pmatrix}
    \sigma^a/2 &\quad&\quad\\ \quad&\quad&\quad
  \end{pmatrix} &
 Q_2^a =
  \begin{pmatrix}
    \quad&\quad&\quad\\ \quad&\quad&-{\sigma^a}^*/2
  \end{pmatrix}.
  \end{align}
 \noindent
 while the generators of the $ U(1)$ are given by
 \begin{equation}
 Y = \diag(1,1,0,-1,-1)/2 \ .
\end{equation}

The electroweak $SU(2)$ is generated by $Q_1^a+Q_2^a$, and is unbroken
by $\Sigma_0$.  The 14 NGBs have definite  quantum numbers under the
electroweak group. We write the NGB matrix as
\begin{equation}
  \label{pgb} 
  \Pi=
  \begin{pmatrix}
    \frac{\eta}{\sqrt{40}}\one
    &\frac{h^T}{2}&\frac\phi{\sqrt2}\\
    \frac{h^*}{{2}}&-\frac{2\eta}{\sqrt{10}}
    &\frac{\tilde{h}}{{2}}\\ \frac{\phi^\dagger}{\sqrt2}&\frac{\tilde{h}^\dagger}{{2}}&
    \frac{\eta}{\sqrt{40}}\one
  \end{pmatrix}
\end{equation}
where $h$ is the Higgs doublet, $h=(h^+, h^0)$, $\tilde{h}=(-h^0, h^+)$, $\eta$ is a real
 neutral field and $\phi$ is an electroweak triplet carrying one unit
 of weak hypercharge, represented as a symmetric two by two matrix.
 We have ignored the three Goldstone bosons that are eaten in the
 Higgsing of $SU(2)^2\times U(1) \to SU(2)\times U(1)$. In the
 following we will also neglect the $\eta$, which is an exact NGB to
 the order in which we are working. In order to avoid phenomenological
 problems from a massless NGB, which, for instance, is constrained by
 rare kaon decays,  we can add small symmetry breaking terms to the
 potential in order to give the $\eta$ a small mass, without
 significantly affecting the discussion of the little Higgs.

The effective theory at the scale $f$ has a tree-level Lagrangian  given
 by
\begin{equation}
  \label{eq:lag}
  {\cal L} = {\cal L}_K + {\cal L}_t + {\cal L}_\psi
\end{equation}
Here ${\cal L}_K$ contains the kinetic terms for all the fields;
${\cal L}_t$ generates the top Yukawa coupling; and ${\cal L}_\psi$
generates the remaining small Yukawa couplings.

The kinetic terms for the fermions and gauge fields are conventional.
The leading two-derivative term for the non-linear sigma model  is
\begin{equation}
  \frac{f^2}{4} \tr\abs{D_\mu\Sigma}^2
\end{equation}
where the covariant derivative of $\Sigma$ is given by
\begin{equation}
  \label{eq:cd}
  D\Sigma = \partial \Sigma - \sum_j \left\{ i g_{j} W_j^a (Q_j^a
   \Sigma + \Sigma Q_j^{aT})  + i g'_ B(Y \Sigma + \Sigma Y^T)\right\}.
\end{equation}
The $g_i, g^\prime$ are the couplings of the $SU(2)^2\times U(1)_y$
  groups.
  
We now turn to  the important  top Yukawa generating sector.  The
largest  radiative corrections to the Higgs potential come from the
top coupling. In the minimal \SM, the top Yukawa coupling leads to a
large, negative, quadratically divergent mass term at one loop.  In
theories  without one loop quadratic divergences, such as softly
broken supersymmetry and little Higgs theories,  the dominant
correction to the quadratic terms in the Higgs potential  come from
the top sector, are negative, and are proportional to the scale of new
physics squared.  We therefore expect new particles associated with
this sector to be lighter than a few TeV.  In
ref. \cite{Arkani-Hamed:2002qy}  the minimal such sector was shown to
contain a pair of colored left handed Weyl fermions
$\tilde{t},\tilde{t}^c$, in addition to the usual third-family weak
doublet $Q = (t,b')$ and weak singlet $\bar U$. In the present model,
we describe a possibility which arises more naturally in composite
model building. We expect the $SU(5)$ symmetry to arise as an
accidental symmetry of the  dynamics of a strongly coupled theory,
analogous to the $SU(3)\times SU(3)$ chiral symmetry of QCD. Composite
fermions  will naturally couple to the composite bosons.  We therefore
introduce new fermions $X,\bar X$ transforming as $(5,3)$ and $(5,
\bar 3)$ respectively under $SU(5)\times SU(3)_{c}$, which gain mass
from the $SU(5)/SO(5)$ symmetry breaking.  These couple to the
$\Sigma$ field in an $SU(5)$ symmetric fashion, and gain mass of order
a TeV. The top mass  will arise by mixing $Q$ and $\bar U$ with
composite quarks of the same quantum numbers, in a manner  similar to
Froggatt-Nielsen models of flavor \cite{Froggatt:1979nt} and the top
see-saw \cite{Dobrescu:1998nm,Chivukula:1998wd}.  Explicitly,  the
fields $X,\bar X$, contain components $\tilde q, \tilde t,p,\bar
p,\bar{\tilde t}, \bar {\tilde q}$, transforming under $SU(3)_c\times
SU(2)' \times SU(2) \times U(1)_y$ as
  
 \begin{center}
 \begin{tabular}{|r|r|cccc|}
\hline & & $SU(3)_c$ & $SU(2)'$ & $SU(2)$ & $U(1)_Y $\\ \hline &
  $p$& 3&1&2&7/6\\ X&$\tilde t$& 3 & 1 & 1 & 2/3\\ &$\tilde q$  &3&2 &
  1 &  1/6 \\ \hline &$\bar {\tilde q}$  &$\bar3$&1 &2  &  -1/6 \\ $\bar
  X$&$\bar{\tilde t}$& $\bar3$ & 1 & 1 & -2/3\\ & $\bar p$&
  $\bar3$&2&1&-7/6\\ \hline
\end{tabular}

\vspace{ 0.2 in}

{ { \large{Charged 2/3 Vector-Like Quark Content}}}

\end{center}
\noindent
We break the $SU(5)$ symmetry only through explicit fermion mass terms
connecting the $Q$ and $\bar T$ to the components of $X, \bar X$ with the
appropriate quantum numbers.  The top Yukawa coupling arises from

\begin{equation}
  \label{topmass1}
  \CL_{t}=\lambda_1 f \bar X \Sigma^{\dagger} X + \lambda_2 f \bar{\tilde
  q} Q + \lambda_3  f \bar T \tilde  t + {\rm h.c.}
  \end{equation}

Because all three terms are needed to entirely break all the symmetry
protecting the little Higgs mass, this form of symmetry breaking is
soft enough to avoid quadratic or logarithmic divergences at one loop,
or quadratic divergences at two loops. Thus at one loop, the largest
radiative corrections to the Higgs potential are insensitive to the UV
and computable in the low energy effective theory.

To see that ${\cal L}_t$ generates a top Yukawa coupling we expand
${\cal L}_t$ to first order in the Higgs $h$:
\begin{equation}
  {\cal L}_t \supset \lambda_1  \bar{\tilde t}{\tilde q} h +
f(\lambda_1  \bar{\tilde t} + \lambda_3 \bar T)\tilde{t}  +f
\bar{\tilde q}( \lambda_1 \tilde q+ \lambda_2 Q) + \cdots\ .
\end{equation}
Clearly $\tilde{t}$ marries the linear combination $(\lambda_1
\bar{\tilde t} + \lambda_3 \bar T)/(\lambda_1^2+\lambda_3^2)^{1/2}$
to become massive,  $\bar{\tilde q}$ marries  the   linear combination
$( \lambda_1 \tilde q+ \lambda_2 Q)/(\lambda_1^2+\lambda_2^2)^{1/2}$,
and $p$ pairs up with $\bar p$.  We can integrate out these heavy
quarks.  The remaining light combinations are $q_3$, the  left handed
 doublet comprised mostly of top and bottom, \begin{equation} q_3 \equiv\frac{(\lambda_2
\tilde q- \lambda_1 Q)}{\sqrt{\lambda_1^2+\lambda_2^2}}\ ,
\end{equation}
and $\bar t$, the left handed  quark which is mostly anti-top,
\begin{equation}
\bar t \equiv \frac{(\lambda_3 \bar{\tilde t}- \lambda_1 \bar
T)}{\sqrt{\lambda_1^2+\lambda_3^2}}\ ,
\end{equation}
with a Yukawa coupling to the little Higgs
\begin{equation}
  \label{topyuk}
  \lambda_t \, h\bar t q_3 +{\rm h.c.} \qquad \text{where} \qquad
  \lambda_t = \frac{\lambda_1  \lambda_2\lambda_3}{\sqrt{\lambda_1^2 +
  \lambda_2^2}\sqrt{\lambda_1^2 + \lambda_3^2}}\ .
\end{equation}

Finally, the interactions in ${\cal L}_\psi$ encode the remaining
Yukawa couplings of the \SM. These couplings are explicitly SU(5)
breaking but small enough so that the 1-loop quadratically divergent
contributions to the Higgs mass they induce are negligible with a
cutoff $\Lambda \sim \text{10 TeV}$.  Note that since there may be 
additional fermions at the cutoff which cancel the anomalies involving 
the broken subgroup  we need only to insist the\SM\ anomalies cancel 
in the effective theory at the TeV scale.

We now turn to a detailed discussion of loop effects in this effective
theory, which give the  Higgs an electroweak symmetry breaking
potential.

\subsection{The Effective Potential and Electroweak Symmetry Breaking}
At tree level the orientation of our $\Sigma$ field is undetermined,
and all the NGBs, including the little Higgs, are massless.  We will
compute the NGB  effective potential at one loop order.  Of course,
our non-renormalizable effective theory is incomplete, and we will need
to add new interactions (counterterms) in order to account for the
cutoff sensitivity introduced by radiative corrections. We follow a
standard chiral Lagrangian analysis,  including all operators
consistent with the symmetries of the theory with  coefficients
assumed to be of the order determined by na\"\i{}ve dimensional
analysis\cite{Manohar:1984md,Cohen:1997rt,Luty:1998fk}, that is, of
similar size to the radiative corrections computed from the lowest
order terms with cutoff $\Lambda=4\pi f$.

The largest corrections  come from the gauge sector, due to  1-loop
quadratic divergences. Remarkably, the one loop quadratically
divergent terms  from the SU(2) gauge interactions do not contribute
to the little Higgs mass squared.  The  gauge divergence is
proportional to
\begin{equation}  \frac{\Lambda^2}{16 \pi^2} \tr M_V^2(\Sigma)
\end{equation}
where $M^2(\Sigma)$ is the gauge boson mass matrix in a background
$\Sigma$. $M_V^2(\Sigma)$ can be read off from the covariant
derivative for $\Sigma$ \eqref{eq:cd}, giving a potential
\begin{equation}
\label{eq:cw}
c g^2_{j} f^4 \sum_a \tr \left[(Q_j^a \Sigma)(Q_j^a \Sigma)^*\right] +
c {g'}^2 f^4 \tr \left[(Y \Sigma)(Y \Sigma)^*\right]
\end{equation}
Here  $c$ is an ${\cal O}(1)$ constant which is sensitive to the UV
physics at the scale $\Lambda$.  Note that at second order in the
gauge couplings and momenta \eqref{eq:cw} is the unique gauge
invariant term transforming properly under the global $SU(5)$
symmetry.  This potential is similar to that generated by
electromagnetic interactions in the pion chiral Lagrangian, which
shift the masses of $\pi^\pm$ from that of the $\pi^0$
\cite{Das:1967it}, by an amount which is quadratically sensitive to
the physics of the GeV scale. In analogy to  the QCD chiral
Lagrangian, we assume that $c$ is positive. This implies that the
gauge interactions prefer the alignment $\Sigma_0$ where the
electroweak group remains unbroken.

To quadratic order in $\phi$ and quartic order in $h$, the potential
from \eqref{eq:cw} is \bea\label{eq:hpot} && c  g_1^2 f^2 \left|
\phi_{ij} - \frac{i}{2f} (h_i h_j + h_j h_i)\right|^2 + c g_2^2
f^2\left|{\phi_{ij} + \frac{i}{2f} (h_i h_j  + h_j h_i)}\right|^2
\cr&&+ c g'^2\left(f^2(2 h^*_i h_i+ 4
\phi^*_{ij}\phi_{ij})-\frac{1}{3} (h^\dagger h)^2\right)\  .  \eea The
term \eqref{eq:hpot} gives the triplet a positive mass squared of
\begin{equation}
m_\phi^2= c(g_1^2+g_2^2 +4{g'}^2) f^2\ .
\end{equation}
 The little Higgs doublet, however, only receives mass at this order
from the $U(1)_Y$ interactions,  because the $SU(2)_{1,2}$ gauge
interactions  each leave  an $SU(3)$ symmetry intact, under which the
little Higgs transforms nonlinearly \cite{Arkani-Hamed:2002qy}.  The
SU(2) interactions do, however, lead to an effective quartic
interaction term in the little Higgs potential, as well as interaction
with the $\phi$ triplet.  After the Higgs triplet is integrated out,
the resulting quartic coupling for the little Higgs is
\begin{equation}
\lambda=c {{ 4 g_1^2 g_2^2 +{
\frac{11}{3}}(g_1^2+g_2^2){g'}^2-\frac{4}{3}{g'}^4}\over
{g_1^2+g_2^2+4 {g'}^2}}
\end{equation}
Note that a miracle seems to occur:  the SU(2) interactions do not lead
to a mass squared for the little Higgs at this order, but do give a
quartic term in the Higgs potential which is of order $ g^2$ when $c$
is of order 1.

The remaining part of the vector boson contribution to the
Coleman-Weinberg potential is
\begin{equation}
\label{eq:cw2}
\frac{3}{64\pi^2}\tr M_V^4(\Sigma)\log
\frac{M_V^2(\Sigma)}{\Lambda^2}\ .
\end{equation} 
This gives a logarithmically enhanced positive Higgs mass squared from
the $SU(2)$ interactions
\begin{equation} 
\label{eq:vector}
\delta m_h^2=\frac{9 g^2 {M'_W}^2}{64\pi^2}
\log\frac{\Lambda^2}{{M'_W}^2}
\end{equation} 
where $M'_W$ is the mass of the heavy $SU(2)$ triplet of gauge bosons.
This contribution can be less than of order 10 times the required
value,  as is needed to avoid more than 10\%  fine-tuning, provided
the $W'$ is lighter than of order 6 TeV.  There is a similar
Coleman-Weinberg potential from the scalar self-interactions in
eq. \ref{eq:cw} which also give logarithmically enhanced positive
contributions to the Higgs mass squared:
\begin{equation}
\label{eq:2}
\delta m_h^2= \frac{\lambda}{16\pi^2}M_\phi^2
\log\frac{\Lambda^2}{M_\phi^2}
\end{equation}
where $M_\phi$ is the triplet scalar mass.

In this theory the top drives  electroweak symmetry breaking, due to
a similar quantum correction to the one which gives radiative electroweak
symmetry breaking in the MSSM. Unlike in the MSSM, however, the one
loop  correction from the top sector has no contribution proportional
to the logarithm of the cutoff, allowing for a somewhat higher new
physics scale.   The quark    loop contribution to the one loop
potential  is
\begin{equation}
\label{topcw}
- 			       \frac{3}{16\pi^2} \tr \left(M_f(\Sigma)
M_f^\dagger(\Sigma)\right)^2\log\frac{M_f(\Sigma)
  M_f^\dagger(\Sigma)}{\Lambda^2}
\end{equation}
where $M_f(\Sigma)$ is the fermion mass matrix in a background
$\Sigma$.  We can neglect the contributions of the light fermions to
this potential, and only consider the effects of the heavy charge 2/3
quarks contained in
$\tilde{t},\bar{\tilde{t}},\tilde{q_t},\bar{\tilde{q_t}},p_t,\bar{p}_t,Q_t$
and $\bar{T}$. Here $Q_t, \tilde{q_t},\bar{\tilde{q_t}},p_t,\bar{p}_t$
denote the charge 2/3 components of the respective weak doublets. 
The charge 2/3 quark mass matrix is

\begin{center}
\begin{tabular}{|r|cccc|}
\hline    & $p_t$ & $\tilde{t}$&$\tilde{q}_t$&$Q_t $\\ \hline
$\bar{p}_t$& $\lambda_1 f\cos^2\theta$& $\lambda_1 f \frac{i
}{\sqrt2}\sin2\theta$&$-\lambda_1 f\sin^2\theta$&0\\
$\bar{\tilde{t}}$&$\lambda_1 f \frac{i }{\sqrt2}\sin2\theta$&
$\lambda_1 f \cos 2\theta$& $\lambda_1 f \frac{i }{\sqrt2}\sin2\theta$
& 0\\ $\bar{\tilde{q_t}}$&$-\lambda_1 f\sin^2\theta$  &$\lambda_1 f
\frac{i }{\sqrt2}\sin2\theta$&$\lambda_1 f\cos^2\theta$ & $\lambda_2
f$ \\ $\bar{T}$&0  &$\lambda_3 f$&0&0 \\ \hline
\end{tabular}

\vspace{ 0.2 in}

{ { \large{Charged 2/3 Quark Mass Matrix}}}
\end{center}

\noindent where $\theta=\langle h\rangle/(\sqrt2 f)$.  Note that
\begin{equation}
\frac{\partial}{\partial \theta}{\rm Tr} M^\dagger M=0\end{equation}
and \begin{equation}\frac{\partial}{\partial \theta}{\rm Tr}
(M^\dagger M)^2=0\end{equation} which guarantees cutoff insensitivity
of the one loop radiative corrections to the Higgs potential from this
sector.  Besides the top which has mass $\lambda_t \langle h\rangle$,
there are three heavy quarks, of mass  \beq M_1&=&\lambda_1 f \cr
M_2&=&\left(a^2 +\frac{ \lambda_t^2\langle h \rangle ^2b^2}{a^2 - b^2}
- \frac{\lambda_t^4\langle h\rangle^4(a^4 - a^2b^2 + b^4)}{(a^2 -
b^2)^3} +\CO(\langle h\rangle ^6)\right)^{1/2}\cr M_3&=&\left(b^2
-\frac{\lambda_t^2 \langle h\rangle ^2a^2 }{a^2 - b^2} +
\frac{\lambda_t^4\langle h \rangle ^4(a^4 - a^2b^2 + b^4)} {(a^2 -
b^2)^3} +\CO(\langle h \rangle ^6)\right)^{1/2} \eeq where \beq
a^2&=&(\lambda_1^2+\lambda_2^2) f^2 \cr
b^2&=&(\lambda_1^2+\lambda_3^2) f^2 \ .\eeq We denote these three
heavy charge 2/3  quarks as the the $t',t'', t'''$, respectively. Note
that if mixing terms of order $\langle h \rangle /f$ are neglected,
these quarks have vector-like \SM\ gauge quantum numbers (3,2,7/6),
(3,1,2/3), and (3,2,1/6) respectively.  Including the top and   $t'',
t'''$ in equation \ref{topcw} gives a contribution to the little Higgs
effective potential  \beq
\label{topx}
\delta V(h)_{\rm eff} =&&-\frac{3\lambda_t^2 h^\dagger h}{8 \pi^2}
\frac{a^2b^2}{a^2-b^2}\log\frac{a^2}{b^2} \cr &&+ \frac{3\lambda_t^4
(h^\dagger h)^2}{16 \pi^2} \left(
\frac{(a^2+b^2)\left((3a^4+3b^4-4b^2a^2)\log\left(\frac{a^2}{b^2}\right)-
(a^4-b^4)\right)} {(a^2-b^2)^3} +2 \log\left( \frac{ab}{h^2}\right)
\right)\cr&&+ \CO(h^6) \  . \eeq  Note that the contribution to the
mass squared is negative while the contribution to the quartic term is
positive, and numerically non-negligible. The negative contribution to
the quadratic term from the top sector is the origin of electroweak
symmetry breaking.   Realistic electroweak symmetry breaking  is
possible  for, {\it e.g. } $M_2\sim M_3\sim 2.8$ TeV, $M_1\sim 2$ TeV,
$c= 1$, $g_1/g_2=5$, $\Lambda\sim10$ TeV and $f= 1$ TeV.  For these
parameters, the  physical Higgs particle mass is  about 480
GeV\footnote{At this point the reader may be concerned about a
discrepancy between such a heavy Higgs and precision electroweak
bounds. Note that a Higgs of up to 500 GeV  can be consistent with
data when there are other nonstandard  corrections
\cite{Chivukula:2000px,Chivukula:2001er}. Note also that a lighter
Higgs may easily be accommodated with smaller value for $c$, and an
additional positive contribution to the Higgs mass squared from some
additional symmetry breaking source.}, and  the quadratic terms in the
Higgs potential from, respectively, gauge loops and fermion loops are
about +(420 GeV)$^2$ and -(550 GeV)$^2$. Note this represents about
40\% cancellation in the quadratic terms, which is not fine tuned.  As
$f$ (and the masses of the heavy particles) is increased, the amount
of fine-tuning necessary to obtain the correct electroweak symmetry
breaking scale will scale as $1/f^2$, and models with $f\gsim 2$ TeV
will typically be more than 10\%  fine-tuned.

\section{UV completion: The little Higgs as a composite Higgs}
\label{comp}
The $SU(5)\to SO(5)$ symmetry breaking pattern  can easily arise from
fermion condensation through technicolor-like strong interactions, as
in an old Composite Higgs model of Dugan, Georgi and Kaplan
\cite{Dugan:1985hq}.  In this section we describe such a UV
completion of the nonlinear sigma model model into a composite Higgs
model. The $SU(5)/SO(5)$ symmetry breaking pattern arises from
condensation of a new set of fermions, called Ultrafermions, which
transform in a real representation of a new gauge group, called
Ultracolor. For concreteness, and because it leads to an elegant
mechanism for giving the top a large mass, we will take Ultracolor to
be an SO(7) gauge group. The generation of fermion masses will require
four fermion operators.   It is conceivable that such
operators might be nearly marginal in a strongly coupled, nearly
conformal theory, in which case a theory with such operators might be UV complete up to very large energies. One might also consider generating them from an extended, Higgsed gauge group, as in extended ultracolor. In the present paper, however,  we will assume the
needed operators are generated by heavy scalar exchange.  Such scalars
are natural in a softly broken supersymmetric theory with a high supersymmetry breaking scale. 

In this section we describe such a  little Higgs  theory from the top down. This theory is UV complete up to a scale as high as the Planck scale. At  high energy we have a
supersymmetric theory with soft supersymmetry breaking at 10
TeV. Although the UV physics is supersymmetric, the supersymmetry is irrelevant for
phenomenology, as all superpartners are beyond direct experimental
reach for the foreseeable future, and are too heavy to lead to
indirect signals such as  flavor changing neutral currents or lepton
flavor violation.

\subsection{Matter content above 10 TeV}
\label{uv}

We begin  our description with a complete list of all the matter superfields in the
theory and their gauge transformations under   $SO(7)\otimes
SU(3)\otimes SU(2)\otimes SU(2)\otimes U(1)$:

\begin{center}
{ {\bf  \large{Chiral Superfields in the theory above 10 TeV}}}

\vspace{ 0.2 in}

\begin{tabular}{|r|c c c  cc|}
\hline & $SO(7)$& $SU(3)_c$ & $SU(2)' $&$SU(2)$&$U(1)_Y $\\ \hline
$L_i$ & 1 & 1 & 1 &2& -1/2 \\ $Q_i$& 1 & 3 & 1&2 & 1/6\\ $\bar U_i$
&1&$\bar 3$ & 1 & 1 & -2/3 \\ $\bar D_i$&1& $\bar 3$ & 1 & 1 & 1/3 \\
$\bar E_i$&1& $1$ & 1 & 1 & 1 \\ 
$\Phi_{\bar 3}$ & 7 & $ \bar 3$ & 1 & 1&-2/3 \\  $\Phi_{ 3}$ & 7 & $
3$ & 1 & 1&2/3 \\ $\Phi_{2'}$&7&1&2&1&-1/2\\ $\Phi_{ 2}$&7&1&1&2&1/2\\
$\Phi_{ 0}$&7&1&1&1&0\\ 
$\bar Y $&1&$\bar3$&1&2&-7/6\\ 
$Y$&1&3&2&1&7/6\\
\hline
\end{tabular}
\smallskip
\end{center}
\noindent
Here $i=1,2,3$ is a generational index. $SU(3)_c$ and $U(1)_Y$ are color and weak hypercharge respectively. Weak isospin is the diagonal subgroup of the two $SU(2)$'s, and $SO(7)$ is the ultracolor group responsible for the fermion condensate which dynamically breaks $SU(5)\rightarrow SO(5)$.   The  fields
$Y$  and $ \bar Y$ are included  to cancel  $SU(2)^2 U(1)$ and $SU(2)'^2 U(1)$
anomalies, so that the theory is free of all  gauge anomalies.

The approximate SU(5) global symmetry of the littlest Higgs nonlinear
sigma model acts on the fields $\Phi_2$, $\Phi_{2'}$, and
$\Phi_0$. The fermion components of these fields will bind to form the composite Higgs.  The $SU(5)$ symmetry is
explicitly broken by the $SU(2)'\times SU(2)\times U(1)_Y$ gauge
interactions and by the superpotential  interactions below.

\subsection{The UV Lagrangian }
The Lagrangian of the theory above the 10 TeV scale contains  the
gauge interactions and the following superpotential interactions:  
\beq \label{UVsuperpot} W=
h_q Q_3 \Phi_{\bar 3} \Phi_2 + h_{\bar u} \bar U_3 \Phi_3\Phi_0+ h'_s
Y \Phi_{\bar 3}\Phi_{2'}+ h_s \bar Y \Phi_3\Phi_{2} + m_3 \Phi_{\bar
3} \Phi_3 + m_0 \Phi_0\Phi_0. \quad \eeq  We leave aside for now the interactions which will generate the light quark and lepton masses. We assume none of the
superpotential  interactions are strong, and that at a scale of 10 TeV
the  SO(7)  gauge theory is near a  strongly coupled superconformal
fixed point.    We then break SUSY softly at around 10 TeV by adding mass
terms \beq  \CL \supset M_i^2 |\phi_i|^2 + (\tilde{m}_0^2 \phi_0^2 +
\tilde{m}_3^2 \phi_3 \phi_{\bar 3} + m_\lambda \lambda \lambda  +
h.c. ).  \eeq   Here $\psi_i$ and $\phi_i$ are the fermion and scalar
components of   the superfield $\Phi_i$ and $\lambda$ is the SO(7)
gaugino.

We begin with a brief description of the sizes of the above symmetry
breaking masses and couplings, delaying a more detailed discussion to
a later section. We assume   $U(1)_R$ symmetry breaking masses of the
scalars and gauginos are small, as is technically natural.   The
couplings $h_q,h_{\bar u}$ need to be of order one to obtain
the top Yukawa. The couplings $h'_s$ and $h_s$ can be of order one
and serve to give mass to the fields $Y$ and $\bar{Y}$.
The supersymmetric $m_3$ mass term will
play an important role in the  dynamics which follows, and is a few
TeV.  The supersymmetry breaking scalar masses  $M_i$, are of order 10
TeV.   The rest of the mass terms are smaller. In order to  protect
the global symmetry of the little Higgs, $m_0$ should be of  order a
GeV, while $\tilde{m_0}$ can be 100 GeV.  The term $\tilde{m}_3$ can
be anywhere between 10 TeV and 0.  The mass of the  gaugino,
$m_\lambda$, should be of order a few TeV.  The smaller masses are
protected by symmetry and thus there are a variety of mechanisms that
could account for their size; we will  content ourselves here with the
observation that they are natural  in the sense of 't Hooft.

We assume that below the scale of the scalar masses, the SO(7) gauge
group confines almost immediately due to its large gauge coupling.
In what follows we assume that the \SM\ squarks, sleptons and gauginos
have a mass of 10-20 TeV or higher (as allowed by naturalness) and
participate neither in the low energy dynamics, nor the phenomenology
of the model.

\subsection{Dynamics and spontaneous symmetry breaking}

We are interested in physics of the low energy composite states.
Such states can only be composites of the fermions $\psi_i$ and the
gaugino $\lambda$.  Ignoring the weak symmetry breaking interactions from the superpotential and weak gauge interactions,
the  approximate global symmetries of the  theory below 10 TeV are an
SU(11) which acts on the 11 fermions in the  fundamental
representation of SO(7) (the $\psi_i$ fields) and an anomaly free
U(1), carried by $\lambda$ as well as the $\psi_i$ fields. There are
no nice solutions to the 'tHooft anomaly matching conditions for
$SU(11)\times U(1)$, so it is reasonable to expect that  part  of the
global symmetry breaks. The most attractive channel is for a
$\lambda\lambda$ condensate, breaking the U(1). It is conceivable that
the SU(11) symmetry would be spontaneously broken to SO(11) by a
$\psi_i\psi_i$ condensate, but it is also possible to match the
'tHooft conditions in a nice way, with composite  spin 1/2 fermions
formed of $\psi_i\psi_j\lambda$, in an antisymmetric tensor of
SU(11). Note that the anomaly of the antisymmetric tensor of the
SU(11) is 7, so  such massless bound states  match the SU(11) anomaly
of the fundamental fermions. It therefore seems plausible, and even
reasonable  by analogy with supersymmetric gauge dynamics, that the
$\psi\psi$ condensate does not form, and the SU(11) remains unbroken
\footnote{The antisymmetric composite fermions do not match discrete
anomalies, where the symmetry is a $Z_2: \psi_i \rightarrow -\psi_i$.
We thus conjecture that this discrete symmetry is broken by an SU(11)
preserving condensate of the form $<\lambda \psi^{11}>$.}.

Note that if there were fewer fermions, the simple composites would
not match the anomalies of the global symmetries.   Furthermore, the
mass term $m_3$ explicitly breaks the SU(11) symmetry to $SU(5)\times
SU(3)$. Some of the composite  fermions  which are massless in order
to match the SU(5) anomalies contain $\psi_3$ and $\psi_{\bar 3}$ as
constituents.  It therefore seems likely that if the mass term $m_3$
becomes large, the SU(5) chiral symmetry will be spontaneously  broken
to SO(5), certainly if $m_3$ were as large as the confinement scale,
$\Lambda$ of 10 TeV this would happen.   Once the symmetry is broken,
the composite fermions will acquire a mass proportional to $m_3$ to
some power.   In particular, we are interested in the masses of the
top partners (composites transforming as $(5,3)$ and $(5,\bar{3}))$.
From \eqref{topmass1} we see that for a top Yukawa of order one, the
symmetry preserving mass of the top partners should be of order $f$,
rather than the larger $4 \pi f$ expected from naive dimensional
analysis in a strongly coupled theory.  However, in our case  the
mass, besides being proportional to the $SU(5)$ chiral symmetry
breaking scale, must also contain the spurion which breaks the $SU(6)$
symmetry of $\psi_3$ and $\psi_{\bar 3}$ to $SU(3)$.  Thus, it should
be proportional to $4\pi f \frac{m_3}{\Lambda}$.  We should therefore
take the explicit symmetry breaking scale $m_3$ to be a few TeV.   At
the same time, we require that the spontaneous symmetry breaking
scale, $4 \pi f$, should be large, at around 10 TeV, or else the top
Yukawa will be too small
\footnote{Alternatively, it also possible to lower the mass of the top
partners by a factor of  order a tenth from their naive $4\pi f$ value
even if the SU(11) symmetry is broken.  It is conceivable that
increasing the size and varying the phases of  the $m_\lambda$, $m_3$,
and $\tilde{m}^2_3$ soft masses (each of which is renormalized by the
others via strong SO(7) interactions) decreases the mass of top
partners.}.

Let us then make an assumption about the dynamics. In strongly coupled
supersymmetric theories, one can study analogous effects  of explicit
symmetry breaking  in determining spontaneous symmetry breaking, and
the scale of spontaneous breaking can be much  larger than the
explicit symmetry breaking. This is because the confined
supersymmetric theories often have a nontrivial moduli space of vacua,
that is, they have massless scalars, and even a small perturbation, if
it gives a light or massless scalar a tadpole,  can drive a big
vev. No such massless scalars are expected for non supersymmetric
theories. However, if the number of flavors of a  non supersymmetric
theory is tuned to some critical number which divides a chiral
symmetry breaking phase from a non chiral symmetry  breaking phase,
then turning on a small mass (below the strong coupling scale) for one
or more flavors should lead to a phase transition.  It is reasonable to assume that this phase
transition is of second order.   A second order transition necessarily
implies a divergent correlation length, {\it i.e.}, a massless
scalar. If, for a near critical number of flavors,  the spectrum
includes a scalar whose mass is zero for a critical value of some mass
parameter, then, even when this mass parameter is not tuned to the
critical value, this scalar will be anomalously light.  For  a theory
with a second order phase transition and  a near critical  number of flavors,
a small mass term results in a large vev.

We now consider an toy example of such dynamics, and write down an
effective potential for the symmetry breaking order parameters.  These
order parameters, $S_{ij}$, have  the quantum numbers of
$\psi_i\psi_j$ bound states.  As we do not expect to have a weakly
coupled description of the dynamics above 10 TeV, the
following effective potential should only be taken as crude estimation
of  the size of symmetry breaking order parameters and not as an accurate
description of the interactions of $\psi_i\psi_j$ bound states.  With
naive dimensional analysis and $SU(11)$ global symmetry as guides, we
expect the first couple of terms  (in an expansion of $S^2/\Lambda^2$)
of the effective potential to take the following form:

\begin{eqnarray}
\CL \supset  a_1 m_3 \frac{\Lambda^2}{4\pi} S_{3\bar{3}} + h.c.  &+&
a_2 \Lambda^2 tr(S^\dagger S) + 4\pi a_3 m_3 S_{3i} S^{\dagger}_{ij}
S_{j\bar{3}} + h.c. \\ \nonumber &+& (4\pi)^2 \frac{a_4}{11}
tr(S^\dagger S S^\dagger S)  + (4\pi)^2 \frac{a'_4}{66} tr(S^\dagger
S)^2 + \cdots\ .
\end{eqnarray}
The first and third terms  come from the $SU(11)$ to $SU(3)\times
SU(5)$ breaking mass term; all other terms are $SU(11)$ symmetric.
The peculiar normalization of the last two terms indicates their
natural size assuming they induce quantum corrections of order one  to
the mass term  when all the $a_i$'s are of order one. As a consequence
of the first term, $S_{3\bar{3}}$ will get a vev of order $a_1 m_3 /(4\pi
a_2)$.

The  order parameter $\Phi$ of $SU(5)$ to $SO(5)$ breaking is the five
by five submatrix of $S_{ij}$ which corresponds to bilinears of
$\psi_2$, $\psi_2'$, and $\psi_0$. The  vev  $S_{3\bar{3}}$ will
induce the following potential for  $\Phi$:
\begin{eqnarray}
\CL \supset a_2 \Lambda^2 tr(\Phi^\dagger \Phi) &+& (4\pi)^2
\frac{2a'_4}{11} \left(\frac{a_1 m_3}{4\pi a_2}\right)^2
tr(\Phi^\dagger \Phi) \\ \nonumber &+& (4\pi)^2 \frac{a_4}{11}
tr(\Phi^\dagger \Phi \Phi^\dagger \Phi)  + (4\pi)^2 \frac{a'_4}{66}
tr(\Phi^\dagger \Phi)^2 \cdots\
\end{eqnarray}
For a second order phase transition to occur, $a'_4$ should be
negative, with $a_4$ positive and slightly larger, so that there is a
net positive quartic term.   For $m_3$ larger  than a certain critical
value, the vev of $\Phi$, $f$, will be about $\Lambda/(4 \pi)$ (assuming
$a_2$ is of order one).
We will take $f$ to be about a TeV to match our little Higgs model.
Consistent with our assumption that a small mass term 
can lead to a sizable vev, we take the parameters $a_i$ such that
the critical value of $m_3$ will also be a few TeV.  For example,
the choice $a_4=-a'_4=a_1=5$ with all other parameters equal to one, 
gives an $m_3$ of 2 TeV.
Below 10 TeV, we thus
have the desired $SU(5)/SO(5)$ Nambu-Goldstone bosons, coupled to
composite fermions  $X_{ij} \sim \psi_i \psi_j \lambda$, which,  as a
consequence of symmetry breaking, are no longer massless.

Let us now describe the masses and interactions of these fermions.
Including the symmetry breaking vevs from above, we find that below 10
TeV, we get the desired $SU(5)$ preserving  interaction of the little
Higgs top sector as well as couplings of the Nambu-Goldstone modes to
the additional fermions:
\begin{eqnarray}
\label{Xcouplings}
\CL \supset b_1 \frac{4\pi m_3}{ \Lambda} f
X_{5\bar{3}}\Sigma^{\dagger} X_{53}  &+& b_2 \frac{(4\pi
f)^2}{\Lambda} tr( \Sigma^{\dagger} X_{55}\Sigma^{\dagger} X_{55}) \\
\nonumber &+&  b_3 \frac{m_3^2}{ \Lambda}
(X_{3\bar{3}}X_{3\bar{3}} + X_{33}X_{\bar{3}\bar{3}}) + h.c. .
\end{eqnarray}
With a $m_3$ about  a TeV, we   have a top Yukawa of order one.  For
our particular choice for the potential, the vev of $S_{3\bar{3}}$ is
similar to $f$, and thus we find that all composite fermions have
comparable masses, parametrically of order $m_3^2/\Lambda$.  This
might seem surprising at first in a vector like theory where mass
inequalities \cite{Weingarten:1983uj,Vafa:1984tf} imply  composites
are at least as heavy as their constituents (violated for $m_3 <<
\Lambda$).  However, recall that the scalar superpartners of the
$\psi_i$ fields, whose masses are of order the strong coupling scale,
do not decouple from the dynamics.  Since the path integral
measure  is then not necessarily positive, the mass inequalities  do not
apply.
   
Having composite fermions in addition to the necessary  top partners,
with masses also of order a TeV,  could be a common feature of
composite little Higgs models.  With the mechanism of being near a
critical number of flavors, and breaking the symmetry which keeps the
fermions degenerate with a small spurion,  all fermion masses are
generically of the same order.  Thus  near a TeV, for similar models
with SU(11) symmetry, there will be neutral color adjoint fermions,
charge $\pm 4/3$  color triplets and anti-triplets,  fermions of charge
$\pm 1$, a weak triplet, weak doublets of charge $\pm 1/2$, and two
singlets.

\subsection{The explicit symmetry breaking interactions}
Let us now estimate the effect of the various symmetry breaking
interactions, making sure that their contributions to the little Higgs
mass are acceptable.  We will use naive dimensional analysis to
approximate the size of operators in the low energy theory.  First,
consider the $M_i^2$ soft masses, for the  five scalars $\phi_2$,
$\phi_0$, and $\phi_2'$.  These break the $SU(5)\times U(1)$  symmetry
to  $SU(2)^2\times U(1)^3$ and so can give mass to the little Higgs.
We  can parameterize them as  \beq {M^2}^i_j = M^2(\delta^i_j +
{G^{(2)}}^i_j + {G^{(2')}}^i_j), \eeq where $M$ is taken to be near
$\Lambda$, while $G^{(2)}$ and $G^{(2')}$ are the mass shifts of
$\phi_2$ and $\phi_2'$ respectively away from the mass of $\phi_0$.
The lowest order contribution to the Higgs mass comes from a U(1)
preserving operator of the form \beq \Lambda^2 f^2
tr(G^{(2)}\Sigma^\dagger G^{(2')}\Sigma).  \eeq Thus there is no
danger as long as the mass splittings are of order 1/100.  This is
quite natural if the scalar masses  are degenerate to begin with and
acquire splittings due to the symmetry breaking gauge interactions. A
possible reason for their initial degeneracy could be that in the UV
the masses of the five scalars are much smaller  than, for example,
the mass of $\phi_3$. Then, when the SO(7) coupling becomes strong,
the other scalar masses will be additively renormalized in an SU(5)
symmetric fashion.  Next, consider the U(1) breaking smaller masses.
Of these the most stringent constraint is on $m_0$ which allows for a
\beq m_0 \frac{\Lambda^2}{4 \pi} f \Sigma_{00} \eeq term.
Consequently $m_0$ must be around a GeV.  A similar operator bounds
$\tilde{m}_0$ to be 100 GeV.  These terms also provide a mass to the
neutral $\eta$ Nambu-Goldstone boson, and so must be nonzero.

As usual, no gauge or Yukawa interaction single handedly breaks the
symmetry protecting the little Higgs, and so a Higgs mass must involve
a combination of at least two separate interactions or an interaction
with a symmetry breaking mass term.  Due to gauge invariance, a gauge
coupling  spurion can only appear in even powers and since each power
comes with a $1/(4 \pi)$, combining  it with any other interaction
(already suppressed by at least a factor of a 100) will lead to a
total $10^{4}$ suppression which is sufficient.  There is no such
constraint on the Yukawa couplings.  Indeed, it is possible to have
terms with each symmetry breaking Yukawa appearing once,  leading to a
mere $1/(4 \pi)^2$ suppression.  However, the Yukawas are charged
under \SM\ quark symmetries, and under an $SU(3)^2$ subgroup of SU(11).
Thus any term containing each Yukawa once, must include \SM\ fields and
an $SU(3)^2 \rightarrow SU(3)$ breaking spurion, such as  \beq
\frac{m_3}{ \Lambda} \frac{h_{\bar u} h_Q}{4\pi} f Q_3
\Sigma_{02} \bar{u}_3,  \eeq  for example (as well as similar terms
with $\tilde{m}^2_3$ and $m_\lambda m_3$).   These provide a small
correction to the top-Higgs interaction and imply a reasonable shift
to the little Higgs mass.

The symmetry breaking terms also lead to new interactions for the
$X_{ij}$ fermions.  Most importantly they provide the necessary SU(5)
breaking masses of \eqref{topmass1}:  \beq \frac{\Lambda}{4\pi} h_Q
Q_3 X_{2\bar 3} + \frac{\Lambda}{4\pi} h_{\bar u} \bar{u}_3 X_{0 3}.
\eeq These have the right size provided that the Yukawa couplings are
of order one.  In addition, there will be contributions that split the
masses of the composite fermions such as
\begin{eqnarray}
&& \frac{(4 \pi f)^2}{\Lambda} tr(G^{(1,2)}\Sigma^\dagger X
\Sigma^\dagger X),  \frac{g^2 f^2}{\Lambda} tr({Q_i^a}^* \Sigma^\dagger X
{Q_i^a}^* \Sigma^\dagger X), \\ \nonumber && \frac{h_Q^2 m_3}{4\pi
 \Lambda} X_{2 \bar 3} {\Sigma^\dagger}^{2 i} X_{i 3},
\frac{h_{\bar u}^2 m_3}{4\pi \Lambda} X_{0 \bar 3}
{\Sigma^\dagger}^{0 i} X_{i 3},\cdots\,
\end{eqnarray}
all of which are one percent corrections.

Thus, except for small modifications, we indeed recover at low
energies the desired little Higgs theory and symmetry breaking
spurions.  We also find additional fermions besides the top  partners
whose masses and interactions violate the global symmetry only
slightly.

\subsection{Obtaining the desired parameters of the UV Lagrangian}
We would like to briefly discuss some important issues concerning the
superpotential and soft masses of the UV Lagrangian as the SO(7)
theory approaches its superconformal fixed point.  First, in order to
obtain the right top sector, all Yukawa couplings in  the
superpotential have to be of order one.  It would therefore be ideal
if the superpotential couplings of (\ref{UVsuperpot}) were marginal.
Near the fixed point, however, the anomalous dimension of the
$\Phi_i\Phi_j$ meson is -3/11 (up to small corrections due to the
weaker interactions).  All the Yukawa interactions are thus slightly
relevant, and so one must assume that the Yukawa couplings were small
in the UV, before the theory began its flow towards the fixed point.
These then become of order one near 10 TeV where the conformal symmetry 
is broken.

The running of the soft scalar masses near the fixed point must also
be addressed.  For each non-anomalous U(1) symmetry, the corresponding
combinations of soft masses (weighed by the U(1) charges of their
respective fields) is invariant under the RG flow.  On the other hand,
various  combinations of soft masses, such as the one corresponding to
an anomalous U(1), flow to zero \cite{Nelson:2001mq,Strassler:2003ht}.  In our case the
relevant non-anomalous U(1)s  are contained in SU(11), while the under
the anomalous U(1) all fields $\Phi_i$ have the same charge.
Consequently, the sum of all scalar masses runs to zero, while the
differences between masses is fixed.  There is thus a danger that some
of the masses squared will become negative, possibly breaking SO(7).
This can be easily avoided if we introduce an additional flavor and
let the strong group be SO(8). Then we can take the scalar mass
squared of this additional flavor to be a bit below the mass of the
rest of the 11 flavors in the UV (a relative factor of two is
sufficient).  As we flow towards the  IR fixed point this scalar will
get a negative mass squared and  break the gauge group back to SO(7).
Meanwhile, the scalars  of our 11 flavors will have positive masses,
of about a factor of  three less than the mass of the extra flavor.
This follows from the assumption that all 11 flavors have similar
masses, and the requirement that the sum of all 12 masses squared
runs to zero. 

\section{Discrete symmetry and dark matter}
\label{dark}

  Any theory
of electroweak symmetry breaking which avoids unacceptable  levels of
proton decay, must preserve baryon and/or lepton number symmetries to a
good approximation.  If both are preserved,   a Lorentz invariant theory
will conserve a $(-1)^{3B+L+2S}$ parity  (where $S$ is the spin),
which we may call dark matter parity\footnote{ Of course, even in theories with baryon or lepton number violation, such as, {\it e.g.} theories with Majorana neutrino masses,  conservation of a discrete subgroup of baryon and lepton number can make dark matter parity  a good symmetry.}.  All \SM\ particles
have  even parity and therefore the lightest parity odd particle (the
LPOP) will be stable against decay.  In supersymmetric theories this
parity is known as R-parity and the LPOP is the lightest superpartner.  
However {\it any} baryon and lepton number conserving theory of TeV physics containing
either fermions which do not have odd $3B+L$ or bosons that do  have odd $3B+L$
has parity odd particles.  The lightest of these is stable. Stable neutral particles at the weak scale with electroweak
interactions  are good candidates  for dark matter\footnote{A loosely related parity, with similar consequences for dark matter,  is the `T parity' which has been proposed to  eliminate tree level corrections to precision electroweak observables \cite{Cheng:2003ju}.} .   The MSSM, for example, has scalar
baryons, scalar leptons, as well as gauginos and Higgsinos
as the non-leptonic and non-baryonic fermions.  Of the fermions,
a mixture of the Higgsinos and weak gauginos, the neutralino 
is a favorite for dark matter.

In our case, the interactions with the \SM\ fields require
the $X_{53}$ fermions to carry  baryon number and so be parity even.
However, the  $X_{55}$, $X_{33}$, and $X_{3 \bar 3}$ fermions are odd
under dark matter parity (as are the heavier $S_{53}$ scalars).
The lightest of these will  be stable. All the $X$ fermions are of comparable mass but gauge and other symmetry breaking interactions should raise the masses of the $X_{33}$, and $X_{3 \bar 3}$ fermions.   A promising dark matter candidate  is one of the neutral components of
$X_{55}$.    All the components of $X_{55}$ are degenerate to within a few percent. Except for the near-degeneracy, the $X_{55}$ fermions are rather similar to the charginos and neutralinos of SUSY
models. The 10 2-component fields of the $X_{55}$  have the quantum
numbers of the bino, wino and the two Higgsino doublets (of opposite
charges).  In addition, there is a charged fermion which is a weak SU(2)
singlet.  All these fermions get masses from the second term in
\eqref{Xcouplings} of about a TeV.  Once the Higgs gets a vev they
  mix at the ten percent level and there will be
additional one percent mass splitting due to both the Higgs and the
above symmetry breaking operators.  The LPOP will thus mix strongly
with its slightly heavier cousins, which will affect estimates of the relic
abundance.  Exploring the possibility of the LPOP in this model being
the dark matter is very interesting and may place further bounds on 
the various parameters.

\section{Precision Electroweak Corrections}
\label{precision}

Precision electroweak observables place important constraints on new physics \cite{Kennedy:1989sn,Holdom:1990tc,Marciano:1990dp,Peskin:1990zt,Golden:1991ig,Kennedy:1990ib,Altarelli:1991zd}.  The precision electroweak corrections of the minimal $SU(5)/SO(5)$ model and several little Higgs variants have been analyzed in refs. \cite{Hewett:2002px,Csaki:2003si,Han:2003wu,Kilian:2003xt}.  For analysis of a structurally similar model, see \cite{Agashe:2003zs}.  The most important precision electroweak corrections come from  dimension 6 operators which arise when integrating out heavy particles at tree level. These are of order $v^2/f^2$, which is parametrically the same as a one loop minimal \SM\ effect. Corrections of order $v^4/f^4$, or a loop factor times $v^2/f^2$, are comparable to two loop \SM\ effects and are mostly smaller than the available experimental precision. An important exception is the one loop correction to the $\rho$ (or T) parameter from the top sector, which is suppressed by both a loop factor and $v^2/f^2$, but is enhanced by color and fermion multiplicity factors,  by relatively large couplings, and a moderately large log. 

In order to determine the dimension 6 operators in the effective theory at tree level, it suffices to write out the  Lagrangian in the limit  where the little Higgs vev is turned off, and find the mass eigenstates at tree level. One can then integrate out  the heavy particles at tree level by solving their equations of motion to order $1/M_{\rm heavy}^2$.  Only the mass and kinetic terms for the heavy fields  and couplings to light particles which are linear in the heavy fields affect the equations of motion to this order. Furthermore, unless there is a heavy-light-light coupling which is  of order $f$, or unless the heavy particle is a fermion, the kinetic terms for the heavy fields may  be ignored.  It is easy to see in our little Higgs theory the operators relevant for precision electroweak corrections are generated from order $f$ heavy-light-light couplings.  These couplings involve the weak isospin  and hypercharge currents, the little Higgs,  and the third generation quarks.  Various corrections by our extension to precision observables such as mass mixing of the weak gauge bosons with additional gauge bosons  proportional to the Higgs vev, mixing of light fermions with new fermions with different electroweak quantum numbers proportional to the Higgs vev, and heavy particle exchange between light particles, may all be accounted for by minor alterations of the results in \cite{Kilian:2003xt}.  

\section{Flavor}
\label{flavor}
Thus far we have only considered the interactions of third generation quarks with
our new SO(7) sector.  Since there will be mixings between the top and and bottom
and new composite fermions, there will be flavor violations for the third generation.  
However, the lighter generations and the leptons, which do not significantly contribute 
to the Higgs mass at one loop, do not need to have interactions which preserve any global symmetry.
We can therefore imagine coupling them indirectly to the SO(7) sector in the UV theory 
in a way which does not introduce flavor violation besides the \SM\
Yukawas themselves. 

For concreteness, here we will discuss a simple way to generate the terms which give rise to the lighter quark and lepton masses in a renormalizable theory. Namely, we  couple the quarks and leptons to  weak doublet supermultiplets in the same way as in the MSSM, but  give this doublet a large ($\sim 10$ TeV)  supersymmetric mass and no significant vev.
That is, we add to the superpotential \ref{UVsuperpot} the terms
\be 
\mu H_u H_d + \lambda^u_{ij} H_u Q_i \bar U_j + \lambda^d_{ij} H_d Q_i \bar D_j +\lambda^e_{ij} H_d L_i \bar E_j  + H_d \Phi_2 \Phi_0 \ ,
\ee
and  soft supersymmetry breaking  scalar mass terms
\be B\mu  H_u H_d + h.c.  +\tilde M_u^2 |H_u|^2 +\tilde M_d^2 |H_d|^2. \ee 
where $\mu, B\mu, \tilde M_{u,d}$ are of order the supersymmetry breaking scale, and $H_{u,d}$
transform under $SU(2)\times U(1)$ as $(2,\pm 1/2)$. 
  At low energies, we will thus get couplings of the 
quarks and leptons to the little Higgs
\beq
y_u^{ij} Q_i f \Sigma_{02} \bar{u}_j + y_d^{ij} Q_i f \Sigma_{02}^* \bar{d}_j + 
y_e^{ij} E_i f \Sigma_{02}^* \bar{e}_j + h.c.
\eeq
with $y^{u,d,e}\propto \lambda^{u,d,e}$.  These terms contain the light quark masses,
as well as the small mixings between the third and other generations.  Note
that only the top has both left handed and right handed components mixing with
the composite fermion states.  Thus in our model it is naturally heavier than the rest
of the fermions. The Yukawa couplings of the down quarks and charged  leptons to the little Higgs  are generically less than of order $\Lambda^2/(4\pi{\tilde M}^2)$.
Additional dangerous operators which violate could violate flavor at low energy, by assumption,  must be
proportional to the only sources of flavor violation, namely the matrices $\lambda^{u,d,e}$, which for light flavors are nearly aligned with the low energy Yukawa coupling matrices.  Similarly, the flavor violation
in the third generation does not significantly affect the first two due to the
tiny mixing angles.   Therefore, the model safely satisfies current constraints on flavor changing neutral 
currents.  

As for the SUSY flavor problem, we can take the masses of 
all the \SM\ superpartners to be 10 TeV or even significantly higher 
\cite{Nelson:2001mq} without affecting
the naturalness of the soft masses of our SO(7) scalars.  They
will thus have very small contributions to flavor violating 
processes.

The only potentially significant flavor violation in our model involves the third generation quarks.
The lower component of the quark doublet $q_3$ is the CKM linear combination of quarks $V_{ti} d_i$, where $i=d,s,b$. 
Thus 
 operator \ref{topfourfermi} can potentially make a contribution to  $B_s$ and $B_d$ mixing which is competitive with the  standard model, while the operator \ref{topop1} gives flavor changing  $Z$ couplings to the   quarks. In the limit where the upper component of $q_3$ is purely top, the phases in the new contribution to $B_s$ and $B_d$ mixing and to the $Z$-$d_i-d^\dagger_j$ coupling are  the same as the one generated by the standard model  top loops. However, if the upper component of $q_3$ is not an exact mass eigenstate, new CP violating phases can appear in $B-\bar B$ mixing and $b$ quark decay amplitudes, at a level which could be as large as standard model loop effects.

\section{Recap and Conclusions}
\label{conclusions}
We have presented a sequence of natural effective field theories, with
no fine-tuning or phenomenological difficulties, describing electroweak
symmetry breaking. The underlying theory is a  supersymmetric theory,
valid to an energy scale which could be as high as the Planck
scale. This theory becomes strongly coupled at some  high scale, above 10
TeV, and approaches an approximate superconformal fixed point.  At 10 TeV, soft supersymmetry breaking drives the theory into a
confining phase, with an unbroken approximate SU(11) chiral symmetry
and several relatively light composite fermions and scalars, with masses of a few TeV.  At 1 TeV,
explicit breaking of some of the SU(11) chiral symmetry  due to mass
terms drives spontaneous breaking of  the remaining chiral SU(5)
symmetry to an SO(5) subgroup, and all composite fermions become
heavy. Most of the resulting pseudo-Goldstone bosons get mass from
explicit symmetry breaking at the TeV scale, or are eaten by TeV mass
gauge bosons. The notable exception is the little Higgs, a doublet which
receives a small, ultraviolet-insensitive negative mass squared from
loops in the top quark mass sector.  Although this Higgs is a
composite particle, it acts like a weakly coupled elementary scalar in
the effective theory, driving electroweak symmetry breaking at a
naturally low scale.

Our model conserves baryon and lepton numbers. We note that  this model, as well as a large class of such beyond the standard model  theories, contains a new stable particle, due to a conserved ``dark matter parity''. All Standard Model particles are even under this parity, however our model  contains a plethora of parity odd particles.  The lightest  parity odd particle, the ``LPOP'', is likely to be a neutral fermion, which may be a good dark matter candidate.

Precision electroweak corrections can provide important constraints on new theories of electroweak symmetry breaking. We give a general effective field theory analysis of the most important precision electroweak corrections in a large class of theories, and discuss the natural parameter regions in which the model gives an acceptable fit to data.

We have described the model and its phenomena in some detail, not because we  expect its particulars to necessarily occur in nature, but rather because we find it an  instructive example of natural electroweak superconductivity, which differs phenomenologically from the oft-studied MSSM, and hence offers different insights into possible experimental clues to search for in our quest for  a solution to the hierarchy problem. This theory provides an example of dynamical electroweak symmetry
breaking, which phenomenologically resembles the minimal standard
model below a TeV. At the TeV scale, it distinguishes itself via
many new  fermions including a good dark matter candidate, a weak triplet of new gauge bosons, and a
scalar triplet. These new particles may be expected to have striking signatures at the LHC. However exploring the underlying strong dynamics, or uncovering supersymmetry, requires much higher energies.

\appendix

\section{Appendix:  Generic Modifications to the T parameter for  theories with additional quarks}
\label{appendix2} 

The corrections to the rho (or T) parameter,  for additional
vector-like, charged 2/3 and -1/3, isospin doublet quarks are given in Lavoura and 
Silva~\cite{Lavoura:1993np}. In little Higgs models, and in general models 
with vector-like fermions, one frequently considers many isospin doublets with various 
charges.  In this section, we generalize Lavoura and Silva's formalism to accommodate weak doublet and singlet 
quarks with arbitrary charges, and show explicitly how decoupling arises.  To begin, we 
write the weak isospin currents.
\begin{equation}
J_{\mathrm{cc}}^\mu = {1 \over 4}\sum_{Q,i,j}{ \bar{\psi}}^Q_i,\gamma^\mu\Biggl[
{V^Q_L}_{ij}(1 - \gamma_5) + {V^Q_R}_{i,j}(1 + \gamma_5)\Biggr]\,\psi^{Q-1}_j \label{chargecurrent}
\end{equation}
\begin{equation}
J_{3}^\mu = {1 \over 4}\sum_{Q,i,j}\bar{\psi^Q}_i\,\gamma^\mu
\Biggl[{W^Q_{L}}_{ij}(1 - \gamma_5) + {W^Q_{R}}_{i,j}(1 + \gamma_5)\Biggr]\,\psi^Q_j
 \Biggr.
\label{neutralcurrent}
\end{equation}
\noindent
Here, ${\psi^Q_{(L,R)}}_j$ represent left and right handed components respectively of the mass eigenstate quarks of charge $Q$. In the standard model, $V_L^{2/3}$ is the CKM matrix. However the 
generalized CKM matrices, $V^Q_L$ and $V^Q_R$, need not be unitary or square, due to the fact that the mass eigenstates may be mixtures of quarks with different weak charges. The matrices $W_{(L,R)}^Q$, which specify the third component of the weak isospin current in terms of mass eigenstates, are square but not necessarily unitary. These   matrices  may be found as follows.

Define matrices ${X^{(u,d),Q}_{(L,R)}}_{ai}$ such that all weak doublet quarks may be written in terms of mass eigenstates as
\be
 {(\tilde u,\tilde d)_{(L,R)}^Q}_a={X^{(u,d),Q}_{(L,R)}}_{ai}{\psi_{(L,R)}^Q}_i\ .
\ee
Here the quarks ${\tilde u}_a^Q$ and ${\tilde d}_a^Q$ are not mass eigenstates, but are members of weak doublets with third component of isospin respectively up and down. Unlike in the standard model, weak isospin up and down quarks  do not necessarily have charges +2/3 and -1/3. The mass eigenstates are defined such that all masses are real and positive.  We continue to label the quarks by electric charge since the mixing matrices can never violate electric charge conservation.
The matrices $X$ satisfy the unitarity conditions
\be
\sum_i {X^{(u,d),Q}_{(L,R)}}_{ai}{X^{(u,d),Q}_{(L,R)}}^\dagger_{ib}=\delta_{ab}
\ee
and
\be
\sum_i {X^{(u),Q}_{(L,R)}}_{ai}{X^{(d),Q}_{(L,R)}}^\dagger_{ib}=0\ .
\ee
Then 
\be
{V^Q_{(L,R)}}_{ij}=\sum_a{{X^{u,Q}_{(L,R)}}}^\dagger_{ia}{X^{d,Q-1}_{(L,R)}}_{aj}
\ee
and 
\be
{W^{Q}_{(L,R)}}_{ij}=\sum_a\left({{X^{u,Q}_{(L,R)}}}^\dagger_{ia}{X^{u,Q}_{(L,R)}}_{aj}-
{{X^{d,Q}_{(L,R)}}}^\dagger_{ia}{X^{d,Q}_{(L,R)}}_{aj}\right)\ee
Note that
\be W^Q_{(L,R)}=V^Q_{(L,R)}{V^Q_{(L,R)}}^\dagger-{V^{Q+1}_{(L,R)}}^\dagger V^{Q+1}_{(L,R)}\  .
\ee

 Now the $T$ parameter is given by
\begin{eqnarray}
\label{tparam}
T &=& {N_c \over 16\pi s^2 c^2}\Biggl\{
\sum_{Q,i,j} \Biggl[(|{V_L}^Q_{ij}|^2 
+ |{V_R}^Q_{ij}|^2)\,\theta_+ (y_{i}, y_{j})
+ 2\mathrm{Re}({V_L}^Q_{ij}{V_{R}}^Q_{ij})\,
\theta_-(y_{i}, y_{j}) \Biggr] \nonumber \\
&-& \sum_{Q, i< j}
\Biggl[(|{W^Q_L}_{ij}|^2 + |{W^Q_R}_{ij}|^2
)\,\theta_+(y_{i}, y_{j}) 
+ 2\mathrm{Re}({W^Q_L}_{ij}{W^Q_R}_{ij}^\dagger)\,
\theta_-(y_{i}, y_{j})\Biggr] \Biggr\}\\
 \nonumber
\end{eqnarray}
\noindent
where subscript latin indices run over the mass eigenstates, and $s^2$ and $c^2$ are 
$\sin^2{\theta_W}$ and $\cos^2{\theta_W}$.  The functions $\theta_+$, $\theta_-$ and 
$y$ are defined as 
\begin{eqnarray}
\theta_+(y_1,y_2) &\equiv& y_1 + y_2 - {2y_1 y_2 \over y_1 - y_2}\ln{{y_1 \over y_2}},\\
\theta_-(y_1,y_2) &\equiv& 2\sqrt{y_1 y_2}\Biggl({y_1 + y_2 \over y_1 - y_2}\ln{{y_1 \over y_2}} 
- 2\Biggr),\\
y_{(1,2)} &\equiv & {m^2_{(1,2)} \over m^2_z}.\end{eqnarray}
\noindent
These formulae can accommodate general additional fermions as long as only weak isospin doublets and singlets are considered, and can easily be modified to allow for other weak representations.

The decoupling of heavy vector-like particles is not immediately obvious from eq. \ref{tparam}.
Decoupling of heavy particles works  when the heavy mass limit does not require strong coupling, and fails when the heavy mass limit requires taking some coupling to be large. Examples of failure of decoupling are the heavy Higgs and/or top mass limit of the Standard Model. 
In the present case, all the new particles can be made arbitrarily heavy without taking any couplings to be large, although one must fine-tune some parameters.

We now demonstrate decoupling of the one loop corrections to the rho parameter from theories such as ours with  heavy  vector-like quarks, whose masses can be made arbitrarily
 large without strong coupling. 

 In a decoupling theory with heavy weak SU(2)  doublets and singlets, the heavy mass eigenstates  are  nearly vector-like weak eigenstates. All   doublets come in nearly degenerate pairs (U,D), with mass splitting no larger than $v^2$. The mixing matrices  $V$ and $W$ are restricted as well. We choose  a basis where the matrices $X$ are approximately rows of the unit matrix. The leading correction   is of order $v/M$, with $M$ the mass scale of the heavy quarks. Terms of  order  $v/M$ are only  possible between weak doublets and singlets. To see the effects of these restrictions, we make the following definitions.

Let $n_{_{Q}}$ be the number of doublets  in the weak basis whose $T_3=+1/2$ components have charge Q. Note that $n_{_{(Q+1)}}$ is then  the number of doublets whose lower components have charge Q.  The total number of fermions with charge Q is 
\be N_Q\equiv n_{_{Q}}+n_{_{(Q+1)}} + \tilde n_{_{Q}}\ , \ee
where $\tilde n_{_{Q}}$ is the number of singlets with charge $Q$. Define
\beq
\delta^Q_{ij}&\equiv&\delta_{i j} , {\rm if} j\le  n_{_{Q}}\cr
&=&0, {\rm otherwise} ,
\eeq
where $\delta_{ij}$ is the usual Kronecker delta, and $i,j$ run from 1 to $N_Q$. 
In the following we will use a basis  where, in the limit   $v/M\rightarrow0$,  $i\le n_{_{(Q+1)}}$ labels quarks which are are  $T_3=-1/2$,   $n_{_{(Q+1)}}<i\le n_{_{(Q+1)}} + n_{_{Q}} $ labels quarks which are  $T_3=+1/2$, and,  for  $i>n_{_{Q}}+n_{_{(Q+1)}}$, the quarks  are  weak singlets. Now
\beq
\label{mixingmatrix}
{V^Q_{L,R}}_{ij}&=& \delta^Q_{\left(i-n_{_{(Q+1)}}\right)\ j} +  \tilde V^Q_{ij} +\CO(v^2/M^2)\\
{W^Q_{L,R}}_{ij}&=&\delta^Q_{(i-n_{_{(Q+1)}})(j-n_{_{(Q+1)}})} -\delta^{Q+1}_{ij}\cr
&&+{\tilde V}^{Q \dagger }_ {\left(i-n_{_{(Q+1)}}\right) \ j}+ {\tilde V}^Q_{i\ \left(j-n_{_{(Q+1)}}\right)}- \sum_k\left({\tilde V}^{(Q+1)}_{ik}\delta^{(Q+1)}_{kj} -\delta^{(Q+1)}_{ik} {\tilde V}^{ (Q+1)\dagger}_{kj}\right)\cr&&+ \CO(v^2/M^2)\ .
\eeq 
Here $\tilde V$ is of order $v/M$. Note that  terms of order $v/M$ transform as doublets under $SU(2)_w$. $SU(2)$ invariance of the effective theory requires that  terms of order $v/M$ must  connect   doublet quarks  with singlet quarks. Therefore
$\tilde V^Q_{ij}=0,$ 
if either  
$i\le n_{_{(Q+1)}}+ n_{_{Q}} $
 and 
 $j\le n_{_{Q}} + n_{Q-1}$,
  or if the inequalities 
  $i> n_{_{(Q+1)}}+ n_{_{Q}} $ and $j>n_{_{Q}} + n_{Q-1}$
   are both satisfied.

 Since mass and weak eigenstates coincide up to angles of order $v/M$, and since weak doublets are nearly degenerate,  quark masses satisfy
\be\label{mass} {\rm if }\    i \le  n_{_{Q}}, \  {\rm then} \ m_{Q\ (i+n_{_{(Q+1)}})}^2=m_{ (Q-1)\ i}^2+ \CO(v^2)\ ,
\ee
where $m^2_{Q\ i}$ labels the mass of the quark flavor $i$ with charge $Q$, and $i= 1,2 \ldots N_Q$. 
Substituting in eqns. \ref{mass} and \ref{mixingmatrix} into eqn. \ref{tparam}, we find that the terms of order 1 in the $V$ matrices give contributions proportional to $\theta_{\pm}(M^2/m_z^2, M^2/m_z^2+ \CO(v^2))$, and are of order $v^4/(M^2 m_z^2)$. The terms proportional to $\tilde V^2$ are proportional to $(v^2/M^2)(\theta_{\pm}(M_b^2, M_a^2/m_z^2+ \CO(v^2))-\theta_{\pm}(M_b^2, M_a^2)$ and are likewise of order $v^4/(M^2 m_z^2)$. Finally, the terms in the mixing matrices which are of order $v^2/M^2$ give even smaller contributions to $T$,  of order $v^6/(M^4 m_z^2)$.

\section{Leading corrections to the T parameter in this model}
The full one loop contribution to the $\rho$, or T parameter in the model receives contributions from all the additional charge 5/3, 2/3  and -1/3 quarks, and is a complicated, unilluminating mess. In order to compute the order $v^2/f^2$ piece, one can use the following  effective field theory  treatment.   Integrate out the heavy quarks at the scale $f$,  reproduce their effects via the  operators \ref{rhoop},  \ref{topcustop}, \ref{topcustop2}, with coefficients computed at tree level for the operators \ref{topcustop}, \ref{topcustop2}, and at one loop order for the operator \ref{rhoop}.  
Under renormalization group scaling from $f$ to $v$, including the effects of the one loop top quark Yukawa coupling, these  operators mix, and so  the operators \ref{topcustop},\ref{topcustop2}  will give a log enhanced additive contribution to the coefficient of the operator \ref{rhoop}.   One then computes the effects of the operator \ref{rhoop} on the $W$ and $Z$ masses at tree level.     

Explicitly, the log enhanced piece turns out to be 
\beq
T=\frac{3}{4\pi s_W^2 c_W^2}\frac{m_t^2}{M_z^2}\frac{v^2}{f^2}
\left(s_2^2c_3^4 \log\frac{b^2}{m_t^2}+2s_3^2c_2^4\log\frac{a^2}{m_t^2}-2s_3^2\log
\frac{(\lambda_1 f)^2}{m_t^2}\right).
\eeq
Here,
\begin{eqnarray} 
s_2 &=& \frac{\lambda_2}{\sqrt{\lambda_1^2+\lambda_2^2}},\  c_2^2=1-s_2^2, \\ \nonumber
s_3 &=& \frac{\lambda_3}{\sqrt{\lambda_1^2+\lambda_3^2}},\  c_3^2=1-s_3^2.
\end{eqnarray}

Numerically, this is gives the dominant correction to T and is of order -0.2 
for $\lambda_1=\lambda_2=\sqrt{6}$,with $\lambda_3=\sqrt{3}$. 
Other non-enhanced pieces will give a somewhat smaller contribution.
\section*{Acknowledgements}

The work of E. Katz,   Jae Yong Lee, and A. Nelson was partially
supported by the DOE under contract DE-FGO3-96-ER40956.  D. Walker is supported by National Science Foundation grant NSF-PHY-9802709. We thank Nima Arkani-Hamed, Andrew Cohen, Aaron Grant, Thomas Gregoire, Wolfgang Kilian, Hitoshi Murayama, and Jay Wacker for useful discussions.

\bibliography{moose}

\bibliographystyle{JHEP}

\end{document}